\documentclass[journal=jctcce,manuscript=article]{achemso}

\usepackage[version=3]{mhchem} 
\usepackage{amssymb,amsmath,mathrsfs,amsbsy,amscd,verbatim}
\usepackage{color,graphicx,float,bm,enumerate}
\usepackage{alltt,listings,tikz}
\usepackage[all]{xy}
\usepackage{amsthm,hyperref}
\newcommand{\drho}{{\delta\rho}} 
\newcommand{\dwrho}{{\delta\widetilde{\rho}}}
\newcommand{\dH}{{\delta H}}
\newcommand{\wH}{{\widetilde{H}}}

\newcommand{\ie}{{\it i.e.}}
\newcommand{\eg}{{\it e.g.}}
\newcommand{\veps}{\varepsilon}

\author{Weiqi Chu}
\email{chu@math.ucla.edu}
\affiliation{Department of Mathematics, University of California, Los Angeles, Los Angeles, CA}
\alsoaffiliation[PennState]{Department of Mathematics, Pennsylvania State University, University Park, PA}

\author{Xiantao Li}
\email{XLi@math.psu.edu}
\affiliation[PennState]{Department of Mathematics, Pennsylvania State University, University Park, PA}

\title[Reduced-order Modeling for Electron Transport]
 {A reduced-order modeling approach for electron transport in molecular junctions}

\abbreviations{IR,NMR,UV}

\begin{document}


\begin{abstract}

To describe non-equilibrium transport processes in a quantum device with infinite baths, we propose to formulate the problems as a reduced-order problem.
Starting with the Liouville-von Neumann  equation for the density-matrix, the reduced-order technique yields a finite system with open boundary conditions. 
We show that with appropriate choices of subspaces, the reduced model can be obtained systematically from the Petrov-Galerkin projection. The self-energy associated with the bath
emerges naturally. The results from the numerical experiments indicate that the reduced models are able to capture both the transient and steady states. 

\end{abstract}

\section{Introduction}

In the past decades, there has been significant progress in the investigation  of molecular electronics and quantum mechanical transport \cite{aviram1989molecular,reed1999molecular,joachim2000electronics}, one emerging issue among which is the modeling of interfaces or junctions between molecular entities \cite{aradhya2013single,cahen2005energetics,cahen2003electron,hwang2009energetics}. The junctions encompass two sections: (i) a molecular core at the nanometer scale that bridges two metallic devices; (ii) the surrounding areas from contacting materials. Notable  examples include quantum dots, quantum wires, and molecule-lead conjunctions. The junctions play an essential role in determining the functionality and properties of the entire device and structure, such as photovoltaic cells \cite{brabec2004organic,bredas2009molecular}, intramolecular vibrational relaxation \cite{poulsen2001path,potter2000transport,everitt1998vibrational,nibbering1991femtosecond}, infrared chromophore spectroscopy, and photochemistry \cite{pshenichnikov1995time,joo1995ultrafast,becker1989femtosecond,lang1999aqueous}. At such a small spatial and temporal scale, modeling the transport properties and processes demands a quantum theory that directly targets the electronic structures.

Such problems have been traditionally treated with the Landauer-B\"{u}ttiker formalism \cite{landauer1970electrical,buttiker1985generalized,buttiker1986four}, which aims at computing the steady-state of a system interacting with two or more macroscopic electrodes, and the non-equilibrium Green's function (NEGF) approach, which, often based on the tight-binding (TB) representation, can naturally incorporate the external potential and predict the steady-state current \cite{cini1980time}. This approach was later extended to the first-principle level \cite{lang1995resistance,taylor2001ab,brandbyge2002density} using the density-functional theory (DFT) \cite{hohenberg1964inhomogeneous,Kohn1965}.

Due to the dynamic nature and the involvement of electron excitations, one natural computational framework for transport problems is the  time-dependent density-functional theory (TDDFT) \cite{kurth2005time,runge1984density,stefanucci2004time,burke2005density,cheng2006simulating}, which extends the DFT to model electron dynamics. This effort was initiated by Stefanucci and Almbladh \cite{stefanucci2004time-prb,stefanucci2004time}, and Kurth et al. \cite{kurth2005time},  where the wave functions are projected into the center and bath regions. An algorithm was developed to propagate the wave functions confined to the center region so that the influence from the bath is taken into account. This is later treated by using the complex absorbing potential (CAP) method \cite{baer2004ab} by Varga \cite{varga2011time}. One computational challenge from this framework is the computation of the initial eigenstates. Kurth et al. \cite{kurth2005time} addressed this issue by diagonalizing the Green's function. However, the normalization is still nontrivial, since the wave functions also have components in the bath regions. Another issue is that the CAP method is usually developed for constant external potentials. For time-dependent scalar potentials, a gauge transformation is usually needed  to express the absorbing boundary condition \cite{antoine2003unconditionally}, and it is not yet clear how this can be implemented within CAP.

Another framework is based on the Liouville-von Neumann  (LvN) equation \cite{sanchez2006molecular,subotnik2009nonequilibrium} to compute the density-matrix operator directly. 
One advantage of the LvN approach is that the initial density-matrix can be obtained quite easily from the Green's function. Therefore diagonalization and normalization are not needed. 
To incorporate the influence of the bath, the LvN equation has been modified by adding a driving term at the contact regions according to the potential bias. This approach was later extended by Zelovich and coworkers  \cite{zelovich2014state,zelovich2015molecule}, which is again motivated by the CAP method. Despite the heuristic derivation  \cite{zelovich2014state}, these methods are still empirical in modeling the electron transport problem.  
In particular, the steady state and transient predicted by the driven LvN equation have not been compared with those from the full model. 

This paper follows the density-matrix-based framework. Rather than using the approach by S\'{a}nchez et al. \cite{sanchez2006molecular}, we derive the open quantum system using the reduced-order techniques that have been widely successful in many engineering applications \cite{bai2002krylov,freund2000krylov,villemagne1987model}. We first formulate the full quantum system as a large-dimensional dynamical system with low-dimensional input and output. This motivates a subspace projection approach, which has been the most robust method in reduced-order modeling \cite{bai2002krylov,freund2000krylov}. In particular, we employ the Petrov-Galerkin projection, a standard tool in numerical computations, \eg, linear systems, eigenvalue problems, matrix equations, and partial differential equations (PDEs)\cite{smith1985numerical,johnson2012numerical,lambert1973computational,morton2005numerical}. 
With appropriate choices of the subspaces, we obtain a reduced LvN equation, modeling an open quantum system where the computational domain only consists of the center and contact regions. 
We illustrate the procedure for a one-dimensional model system, as a first step to treat more realistic systems. The numerical results have shown that the reduced LvN equations can capture both the transient and the steady state solutions. 

The rest of the paper is organized as follows. In Sec. \ref{sec: 2}, we provide a detailed account of our methodology, including the mathematical framework and the derivation of the reduced models. In Sec. \ref{sec: results}, we present results from some numerical experiments to examine the effectiveness of the derived models. Sec. \ref{sec: summary} summarizes the methodology and provides an outlook of future works.

\section{Methods and algorithms}\label{sec: 2}
\subsection{The density-matrix formulation}
\begin{figure}[H]
	\includegraphics[width=0.8\textwidth]{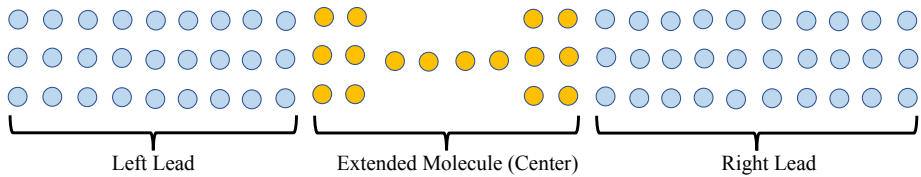}
	\caption{(Color online) Schematic representation of a two semi-infinite lead junction model consisting of two semi-infinite leads: left lead (L), right lead (R), and an extended molecule (C) in the center.}
	\label{fig: schematic1}
\end{figure}

Following the conventions from existing literature \cite{brandbyge2002density,kurth2005time,zelovich2014state,zelovich2015molecule}, we consider a molecular junction, where a molecule is connected to two semi-infinite leads. 
More specifically,  the physical domain for the entire system is denoted by $\Omega$,  divided into three parts, $\Omega_{L}, \Omega_{C}$, and $\Omega_{R}$ representing respectively the left lead, the center region, and the right lead, as illustrated in Figure \ref{fig: schematic1}.

We will start with the LvN equation, which for molecular conduction problems,  has been proposed and implemented in a series of papers. \cite{sanchez2006molecular,subotnik2009nonequilibrium,zelovich2014state,zelovich2015molecule}.  The LvN equation governs the dynamics of the density-matrix operator $\hat{\rho}$,  which can be connected to the wave functions (\eg, the Kohn-Sham orbitals) as follows,
\begin{equation}
\hat{\rho}(\bm r,\bm r',t) = \sum_{j} n_{j} \hat{\psi}_{j}(\bm r,t)\hat{\psi}_{j}(\bm r',t)^{*},
\end{equation}
with $n_{j}$ being the occupation numbers.  The equation  can be derived from a time-dependent Schr\"{o}dinger equation (TDSE), and for the entire system $\Omega$, it can be written as,
\begin{equation}\label{eq: LvN}
	i {\partial_t} \hat{\rho}(t) = \hat{H}(t)\hat{\rho}(t) - \hat{\rho}(t)\hat{H}(t)=[\hat{H}(t),\hat{\rho}(t)].
\end{equation}

Here the bracket is the usual quantum commutator, which we will generalize as follows,
\begin{equation}\label{eq: comm}
	[\hat{A},\hat{B}] := \hat{A}^{*}\hat{B} - \hat{B}^{*}\hat{A}.
\end{equation}
Here $A^*$ denotes the conjugate transpose (or Hermitian transpose of $A$). Notice that with this generalization,   $A$ or $B$ can be non-Hermitian.

Our goal is to derive an {\it open} quantum system for the density-matrix at the center region $\Omega_C$, where the influence from the leads is implicitly incorporated. 
For convenience, we first assume that the entire system \eqref{eq: LvN} has been appropriately discretized in $\Omega$ so that $\rho(\bm r,\bm r',t)$ is a matrix defined at certain grid points, here denoted by $\Omega_{\Delta}$ with $\Delta$ indicating the grid size. Namely,  $\rho(\bm r, \bm r', t)$ is the density-matrix with $\bm r, \bm r' \in \Omega_{\Delta}$. This can be obtained by using a finite-difference scheme, especially in  real-space methods \cite{beck2000real}.  As a result, one arrives at a matrix-valued infinite-dimensional system, and hence we will drop the $\hat{\quad}$ notation from now on.  A similar system can also be obtained using the TB approximation, where the wave functions are projected to atomic-centered orbitals, in which case, the LvN equation would contain the overlap matrix on the left hand side when the basis functions are not orthogonal. \cite{zelovich2015molecule,sankey1989ab} However, it would not affect our following reduction method.

Following the setup by Cini \cite{cini1980time}, we treat the problem as an initial value problem (IVP), starting with an initial density $\rho_{0}=f_{eq}(\mu-H_{0})$ as an equilibrium density at $t=0$. 
Such setup is particularly amenable for numerical computations. While it is challenging to compute the wave function in a subdomain, which in general requires solving nonlinear eigenvalue problems and normalization \cite{Inglesfield1981}, efficient algorithms are available to calculate the density-matrix in a sub-domain \cite{KellyCar92,lin2011selinv,williams1982green}. These algorithms take advantage of the relation between the density-matrix and the Green's function,
 \begin{equation}\label{eq: g-to-rho}
   \rho  = \frac{1}{2\pi i} \oint_C G(z) dz, \quad G(z)= (zI - H)^{-1},
\end{equation}
where the contour encloses all the occupied states. The restrictions of the density-matrix to a finite subdomain can be obtained by $E^* \rho E$, where the operator $E$, with proper arrangement, can be written simply as $E^*=[I, \quad 0],$ with the identity operator $I$ corresponding to the subdomain and the zero matrix corresponding to the exterior (bath). This observation, together with \eqref{eq: g-to-rho}, reduces the problem to the computation of the following expression that we have slightly generalized the linear algebraic system to,
\begin{equation}\label{eq: reduced-G}
  [ \times \quad 0] (zI - H)^{-1} \left[
  \begin{array}{c}
  \times \\
  0 \end{array}\right], 
\end{equation}  
where the left and right vectors have finite supports. 
Although this amounts to solving an infinite-dimensional linear system, a finite number of unknowns are needed due to the multiplication by the sparse vector on the left and right. For one-dimensional (or quasi one-dimensional) systems, an iterative scheme can be used \cite{godfrin1991method,pecchia2008non} to invert the block tri-diagonal matrix. For multi-dimensional problems, a discrete boundary element method \cite{Li2012} can be used \cite{li2016pexsi}.  We will refer to these algorithms in general as  {\it selective inversion }\cite{lin2011selinv}.

Although our model works with the density-matrix, our primary interest is in the electric current induced by a time-dependent external potential that is switched on at $t=0_+$.
Similar to the theory of linear response \cite{gross1985local,dobson1997time,gunnarsson1976exchange}, we consider $H(t)$ as a deviation from its initial value $H_{0}$ and write $H(t)=H_{0}+\dH(t)$ with $\dH(t)$ being the applied potential from the leads. The response of the system due to the external potential could be represented in terms of the perturbed density,
\begin{equation}
 	\drho(t) :=\rho(t) - \rho_{0}, \quad \drho(0)=0,
\end{equation}
which satisfies a response equation,
\begin{equation}\label{eq: drhoeq}
 	i\frac{d}{dt}\drho(t)= [H(t),\drho(t)] + \Theta(t).
\end{equation}
Here $\Theta(t) = [\dH(t),\rho_{0}]$ is a non-homogeneous term that incorporates the influence from the external potential.

As is customary \cite{kurth2005time,li2019absorbing,sanchez2006molecular,zelovich2014state}, we neglect the direct coupling between the two leads and partition the density-matrix and the Hamiltonian operator in accordance with the partition of the domain indicated in Figure \ref{fig: schematic1}. In this case, Eq \eqref{eq: drhoeq} translates to
\begin{equation}\label{eq: blockLvN}
i\frac{d}{dt} 
\left(
\begin{array}{ccc}
\drho_{LL} & \drho_{LC} & \drho_{LR} \\
\drho_{CL} & \drho_{CC} & \drho_{CR} \\
\drho_{RL} & \drho_{RC} & \drho_{RR} \\
\end{array}
\right) 
=
\left[
\left(
\begin{array}{ccc}
H_{LL} & H_{LC} & 0 \\
H_{CL} & H_{CC} & H_{CR} \\
0 & H_{RC} & H_{RR} \\
\end{array}
\right), 
\left(
\begin{array}{ccc}
\drho_{LL} & \drho_{LC} & \drho_{LR} \\
\drho_{CL} & \drho_{CC} & \drho_{CR} \\
\drho_{RL} & \drho_{RC} & \drho_{RR} \\
\end{array}
\right) 
\right] + \Theta.
\end{equation}

We are interested in the case when $\delta H$ corresponds to scalar potentials in the leads, given by $ U_L(t)$ and $ U_R(t).$ Then the matrix function $\Theta(t)$ can be
written as,
\begin{equation}\label{eq: Theta(t)}
\Theta(t) = \left[\begin{array}{ccc} 0  &  U_L(t) \rho_{LC}(0)  &  ( U_L(t) - U_R(t)) \rho_{LR}(0) \\
- U_L(t) \rho_{CL}(0) & 0 & - U_R(t) \rho_{CR}(0) \\
- ( U_L(t) - U_R(t)) \rho_{RL}(0) &   U_R(t) \rho_{RC}(0) & 0\end{array}\right].
\end{equation}

In practice, to mimic the infinite leads, one has to pick much larger regions $\Omega_{L/R}$ to prevent the finite size effect, \eg, a recurrence.  
This makes a direct implementation   using Eq \eqref{eq: blockLvN} impractical and requires model reduction tools to reduce the complexity of the full problem. 

\medskip
There are six unknown blocks in the density-matrix $\delta\rho:$ the blocks $\delta\rho_{LL}$ (and  $\delta\rho_{RR}$) are semi-infinite, and this is where an appropriate reduction is needed.
It suffices to illustrate the reduction of the degrees of freedom in the left bath. A direct computation yields
\begin{equation}\label{eq: leftLvN}
	i\frac{d}{dt} \delta \rho_{LL}(t) = [H_{LL}(t),\delta\rho_{LL}(t)] + F_{L}(t),
\end{equation}
where $H_{LL}(t)=H_{LL}(0)+\dH_{LL}(t)$ and $\dH_{LL}(t)$ is the external potential imposed on the left lead. $F_{L}(t)$ represents the influence from  the interior and can be extracted from \eqref{eq: blockLvN},
\begin{equation}\label{eq: UL}
	F_{L}(t)=H_{LC}\drho_{CL}(t) - \drho_{LC}(t)H_{CL}+{\Theta_{LL}(t)}.
\end{equation}

Now our key observation is that Eqs \eqref{eq: leftLvN} and \eqref{eq: UL} constitute an infinite-dimensional control problem with control variables $\delta \rho_{CL}$ and output $\drho_{LL}$. In practice, only the entries in  $\drho_{LL}$ near the interface (between $\Omega_L$ and $\Omega_C$) are needed. Such a large-dimensional dynamical system with low-dimensional input and output can be effectively treated by using the reduced-order techniques \cite{bai2002krylov,freund2000krylov,gugercin2013model,lucia2004reduced,nayfeh2005reduced}.

\subsection{General Petrov-Galerkin projection methods}
Motivated by the development of reduced-order modeling techniques 
\cite{gugercin2013model,lucia2004reduced,decoster1976comparative} that have been widely used in control problems \cite{villemagne1987model}, circuit simulation \cite{freund2000krylov}, and microelectromechanical systems \cite{nayfeh2005reduced}, etc., we propose a Petrov-Galerkin projection approach to derive a reduced model from the infinite-dimensional LvN Eq \eqref{eq: leftLvN}. The objective is to provide a reduced dynamics for the device region that captures both the transient and the steady state. 

The first ingredient is to pick an appropriate subspace where the approximate solution is sought. 
To start with, we pick an $n$-dimensional subspace $\mathcal{V}_{L}$ spanned by a group of basis functions $\{\varphi_{i}\}_{i=1}^{n}$. The subspace can be expressed in a matrix form as $V_{L}= [\varphi_1, \varphi_2, \cdots, \varphi_n]$:  $\mathcal{V}_{L}=\text{Range}(V_L).$
Throughout this paper, we will not distinguish a subspace $\mathcal{V}_{L}$ and its matrix representation $V_L$.

\begin{figure}
\begin{tikzpicture}[scale = 0.5]
	\draw (-3,0) -- (13,0);
	\foreach \i in {1,6,9,10,11,12}
	{
        		\draw[thick,dashed]  (\i-1,0) -- (\i,4);
        		\draw[thick,dashed] (\i+1,0) -- (\i,4);
	}
	\foreach \i in {-2,...,12}
		\draw[thick] (\i,0) -- (\i,0.3);

	\foreach \i in {4.5}
	{
		\draw (12, \i) node  {$\varphi_1$};
		\draw (11, \i) node  {$\varphi_2$};
		\draw (10, \i) node  {$\varphi_3$};
		\draw (9, \i) node  {$\varphi_4$};
		\draw (6, \i) node  {$\varphi_5$};
		\draw (3.5, \i) node  {$\cdots$};
		\draw (1, \i) node  {$\varphi_n$};
	}
	\draw (5,-0.8) node {$\Omega_{L}$};
\end{tikzpicture}
\caption{A diagram of hat functions on $\Omega_{L}$ that span a subspace $\mathcal{V}_{L}$ with dimension $n$.}
\label{fig: spaceV}
\end{figure}
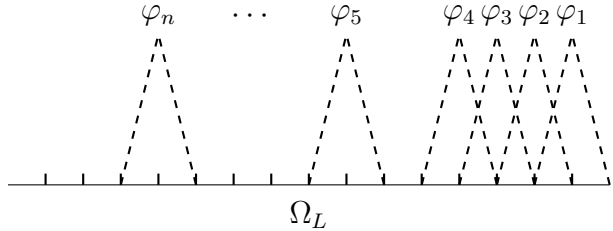

In practice, the basis functions can be standard hat functions  centered at certain grid points, as shown in Figure \ref{fig: spaceV}, or Gaussian-like functions that mimic atomic orbitals.

With the subspace set up,  one can seek a low-rank approximation of $\drho_{LL}(t)$ as $\dwrho_{LL}$ in the following form, 
\begin{equation}\label{eq: g1}
	 {\dwrho}_{LL}(t) := V_{L} D_{LL}(t) V_{L}^{*},
\end{equation}
where  the $n\times n$ matrix $D_{LL}(t)$ represents the nodal values.  This representation automatically guarantees that the resulting density-matrix is Hermitian and
semi positive-definite, as long as $D_{LL}$ has those properties. The residual error from this approximation can be directly deduced from the LvN equation \eqref{eq: leftLvN} by subtraction,
\begin{equation}
   {\mathcal{E}}(D_{LL},t) = iV_L \frac{d}{dt} D_{LL}(t) V_L^* - [H(t), V_{L} D_{LL}(t) V_{L}^{*}] -F_L(t).
\end{equation}
The second ingredient to determine $D_{LL}$ is by projecting the residual error to the orthogonal complement of a test subspace, $\mathcal{W}_L$, spanned by the columns of $W_{L}$, that is 
\begin{equation}\label{eq: g3}
W^{*}_{L}\mathcal{E}(D_{LL})W_{L}=0.
\end{equation}

This yields a finite-dimensional system, and the reduction procedure described above is known in general as the Petrov-Galerkin projection, which has been a classical numerical method in the solutions of differential equations \cite{larsson2008partial},  order-reduction problems \cite{bai2002krylov,freund2000krylov}, and matrix equations \cite{jaimoukha1994krylov,jbilou2006projection}.

The reduced equation from the Petrov-Galerkin projection Eqs \eqref{eq: g1}  to \eqref{eq: g3}  can be written as,
\begin{equation}\label{eq: reducedleftLvN}
	i\frac{d}{dt} D_{LL}(t) = [\wH_{LL}M_{L}, D_{LL}] - \widetilde{F}_{L}(t),
\end{equation}
where the matrices are given by
\begin{equation}
\begin{aligned}
M_{L}&=\left(V_{L}^{*}W_{L}\right)^{-1},\\
\wH_{LL}(t)&=V_{L}^{*}H_{LL}(t)W_{L},\\
\widetilde{F}_{L}(t)&=M_{L}^{*}W_{L}^{*}F_{L}(t)W_{L}M_{L}.
\end{aligned}
\end{equation}
Notice that in \eqref{eq: reducedleftLvN} we have used the generalized notation of commutators \eqref{eq: comm}.
At this point, we will keep the subspaces spanned by $V_L$ and $W_L$ at the abstract level, and the specific choices will be discussed in the next section.

The same model reduction procedure can be applied to the right lead and it yields a similar finite-dimensional equation,
\begin{equation}\label{eq: reducedrightLvN}
	i\frac{d}{dt} D_{RR}(t) = [ \wH_{RR}M_{R}, D_{RR}] - \widetilde{F}_{R}(t).
\end{equation}
Eqs  \eqref{eq: reducedleftLvN} and \eqref{eq: reducedrightLvN} are related by the non-homogeneous terms $\widetilde{F}_{\alpha}(t), \alpha={L,R}$ that involve the evolution of $\drho_{C\alpha}$ and their Hermitian transpose. 

In the center region, no reduction is needed and we will retain this part of Eq  \eqref{eq: blockLvN}. Therefore, we can construct a Petrov-Galerkin projection for the {\it entire} system, by {\it gluing } the subspaces as follows,
 
\begin{equation}\label{eq: VW}
V = \left[ 
	\begin{array}{ccc}
	  V_{L} &  0   &  0 \\
	   0   &  I_{n_{C}}  &  0     \\
	   0   &    0      &  V_{R} \\
	\end{array}
	\right], \quad
W = \left[ 
	\begin{array}{ccc}
	  W_{L} &  0   &  0 \\
	   0   &  I_{n_{C}}  &  0     \\
	   0   &    0      & W_{R} \\
	\end{array}
	\right].
\end{equation}
We seek an approximate solution 
\begin{equation}\label{eq: delrho}
\drho (t) \approx \dwrho(t) := VD(t)V^{*},
\end{equation}
 for the projected dynamics of Eq \eqref{eq: drhoeq}, such that,
\begin{equation}
	i\frac{d}{dt} W^{*}\dwrho(t)W = W^{*} \big( [H(t),\dwrho(t)] + \Theta(t) \big) W.
\end{equation}

Direct computations yield,
\begin{equation}\label{eq: Dequation}
	i\frac{d}{dt} D(t) = [{H}_\text{eff},D] + \widetilde{\Theta}(t),
\end{equation}
where ${H}_\text{eff}$ is the reduced {Hamiltonian},
\begin{equation}\label{eq: tildeH}
	{H}_\text{eff} = 
	\left[ 
		\begin{array}{ccc}
		V_{L}^{*}H_{LL}W_{L}\left( V^{*}_{L}W_{L}\right)^{-1}   &V^{*}_{L}H_{LC} &     0   \\
		H_{CL}W_{L}\left( V^{*}_{L}W_{L}\right)^{-1}   &  H_{CC}      &  H_{CR}W_{R}\left( V^{*}_{R}W_{R} \right)^{-1} \\
		0    &  V^{*}_{R}H_{RC}    &  V^{*}_{R}H_{RR}W_{R}\left( V^{*}_{R}W_{R} \right)^{-1} \\
 		\end{array}
	\right],
\end{equation}
and $\widetilde{\Theta}(t)$ is given by
\begin{equation}
 \widetilde{\Theta}(t) = M^{*}W^{*} \Theta(t) WM = M^{*}W^{*}[\dH(t), \rho_{0}]WM.
\end{equation}
Here the matrix $M$ is block-diagonal,
\begin{equation}\label{eq: M}
M = \left[ 
	\begin{array}{ccc}
	 \left( V^{*}_{L}W_{L}\right)^{-1} &	0	&  0 \\
	 		0	&     I_{n_{C}} 	&    0 \\
			0	&      0		&    \left( V^{*}_{R}W_{R}\right)^{-1} \\
	\end{array}
\right].
\end{equation}

It is  worthwhile to point out that the subspaces can also be time-dependent. This offers the flexibility to pick subspaces that evolve in time.  
It should also be emphasized that our discussions regarding the Petrov-Galerkin projection is suitable for general cases and not limited to one-dimensional junction models, \ie, the typical lead-molecule-lead structures. With appropriate domain decomposition, it can be applied to high-dimensional systems with more general device structures.

\subsection{The selection of the subspaces}
\begin{figure}[H]
	\includegraphics[width=0.7\textwidth]{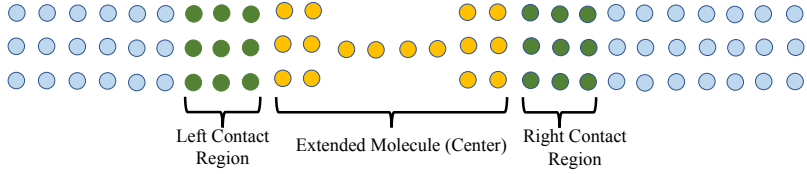}
	\caption{(Color online) A schematic representation of junction model with contact regions in green.}
	\label{fig: schematic2}
\end{figure}

In this section, we discuss specific choices of the subspaces in the Galerkin-Petrov projection. Without loss of generality, we again start by considering the left lead $\Omega_{L}$. Let $\Omega_{\Gamma_{\!L}} \subset \Omega_{L}$ be a subdomain in the left lead that is adjacent to the center region, as shown in Figure \ref{fig: schematic2}. $\Omega_{\Gamma_{\!L}}$ and $\Omega_{\Gamma_{\!R}}$ are often referred to as contact regions that have direct coupling  with the interior \cite{williams1982green,do2014non}. In our case, we pick $\Omega_{\Gamma_{\!R}}$ in such a way that the remaining component in the lead has no coupling with the center region, \ie, $H_{i,j}=0$ for $i \in \Omega_C$ amd $j \in \Omega_{R}-\Omega_{\Gamma_{\!R}}$. This imposes a lower bound on the size of the contact region.

In reduced-order modeling problems, the subspaces are often chosen based on how the input/control variables enter the large-dimensional system, \eg, see the review papers \cite{bai2002krylov,freund2000krylov}.  In our setting, we consider the dynamics in the left lead, given the density-matrix in the contact region.
So we pick the basis $V_L$ so that  $V^{*}_{L}$  acts as a restriction operator from $\Omega_{L}$ to $\Omega_{\Gamma_{\!L}}$,
\begin{equation}\label{eq: VL}
	V^{*}_{L} = [0, I_{n_{\Gamma,L}}],
\end{equation}
where $I_{n_{\Gamma,L}}$ is an identity matrix with the dimension $n_{\Gamma,L}$ being the number of grid points in $\Omega_{\Gamma_{\!L}}$.

The same procedure can be applied to the other lead region. When the subspaces are combined (cf. Eq  \eqref{eq: VW}), we have, 
\begin{equation}\label{eq: V'}
V = \left[ 
	\begin{array}{ccc}
	    0 & 0 & 0,\\
	   I_{n_{\Gamma,L}} &  0   &  0 \\
	   0   &  I_{n_{C}}  &  0     \\
	   0   &    0      &   I_{n_{\Gamma,R}} \\
	   0 & 0 & 0
	\end{array}
	\right].
\end{equation}	
The entire density-matrix is approximated as in \eqref{eq: delrho}. 
 It is now clear that  $V$ is a restriction operator to an extended center domain, $ \Omega_{\widetilde{C}} = \Omega_{\Gamma_{\!L}}\cup \Omega_{C} \cup \Omega_{\Gamma_{\!R}}$. Consequently,  $D$ in Eq \eqref{eq: Dequation} becomes the density-matrix in $\widetilde{C}$,
\begin{equation}
D(t) = \dwrho(t)|_{\Omega_{\widetilde{C}}\times\Omega_{\widetilde{C}}}.
\end{equation}

It remains to choose the subspaces $W_{L/R}$. Motivated by the  Green's function   approach for quantum transport \cite{kadanoff1962quantum,caroli1971direct,datta2005quantum}, we consider the test space,
\begin{equation}\label{eq: WL}
	W_{L}(\varepsilon) = (\varepsilon I - H_{LL})^{-1}V_{L},
\end{equation}
where $\varepsilon \in \mathbb{C}$  is in the resolvent space of the Hamiltonian $H_{LL}$. We require that $\text{Im} \big(\varepsilon\big) <0 $ to ensure the stability of the reduced models. In this case, it corresponds to the advanced Green's function as the imaginary part of $\varepsilon$ goes to zero,  
\begin{equation}
  \lim_{\text{Im}(\varepsilon)\rightarrow 0_{-}} W_{L}(\varepsilon)  =G_L^{A}(\varepsilon)V_L.
\end{equation}
The selection of $W_{R}$ is similar. Intuitively, the subspace $W$ obtained this way represents the solution of the corresponding TDSE with initial conditions supported in the  extended device region $\widetilde{C}$. Combining the subspaces $W_L$ and $W_R$, we have
\begin{equation}\label{eq: W'}
W = \left[ 
\begin{array}{cll}
  W_L  & 
  \begin{matrix}
    0 \\
    0\\
    \end{matrix} &
      \begin{matrix}
    0 \\
    0\\
    \end{matrix} 
      \\
    0&  I_{n_{C}}   &0 \\
  \begin{matrix}
    0 \\
    0\\
    \end{matrix} &
      \begin{matrix}
    0 \\
    0\\
    \end{matrix} &
    W_R \\
    \end{array}   \right].
 \end{equation}
We notice in passing that unlike the basis $V_L$ amd $V_R$, the basis $W_L$  and $W_R$ do not   have compact support.

We now examine the specific form of the reduced model \eqref{eq: Dequation}. With the specific choices of the subspaces (Eqs  \eqref{eq: V'} and \eqref{eq: W'}),  one can simplify the matrix  $M$ in Eq  \eqref{eq:  M} as follows, 
\begin{equation}\label{eq: mass}
	M_{LL} = \left(V^{*}_{L}W_{L}\right)^{-1} = \varepsilon I - H_{\Gamma_{\!L},\Gamma_{\!L}}(t) - \Sigma_{L}(t,\varepsilon)
	            =: \varepsilon I - H_{\text{eff},L}(t,\varepsilon),
\end{equation}
and similarly,
\begin{equation}
  M_{RR} =  \varepsilon I - H_{\text{eff},R}(t,\varepsilon).
\end{equation}
Here $\Sigma_{\alpha}$ is the self energy \cite{brandbyge2002density,popov2004tunnel,danielewicz1984quantum,xue2002first} contributed by the left ($\alpha=L$) or right ($\alpha=R$) lead, 
\begin{equation}\label{eq: selfenergyL}
	\Sigma_{\alpha}(t,\varepsilon)=H_{\Gamma_{\!\alpha},\alpha} (\varepsilon I- H_{\alpha,\alpha}(t))^{-1} H_{\alpha,\Gamma_{\!\alpha}},
\end{equation}
and $H_{\text{eff},\alpha}$ is the effective Hamiltonian associated with $\Omega_{\Gamma_{\alpha}}$ \cite{meier1999non},
\begin{equation}
H_{\text{eff},\alpha}(t,\varepsilon)=H_{\Gamma_{\!\alpha},\Gamma_{\!\alpha}}(t)+\Sigma_{\alpha}(t,\varepsilon).
\end{equation}

 Overall, the effective Hamiltonian $H_\text{eff}$ in \eqref{eq: Dequation} is simplified to,
\begin{equation}
H_\text{eff}(t):=H_c(t)+\Sigma(t,\varepsilon),
\end{equation}
 where ${H}_{c}$ is the Hamiltonian restricted in the extended center region $\Omega_{\widetilde{C}}$,
\begin{equation}
\begin{aligned}
	H_c(t) :=  H(t) \large|_{\Omega_{\widetilde{C}}\times\Omega_{\widetilde{C}}},
\end{aligned}
\end{equation}
and
$\Sigma$ is a block-wise diagonal matrix that incorporates  the self-energies of two leads,
\begin{equation}
	\Sigma(t,\varepsilon) = \left[
		\begin{array}{ccc}
		\Sigma_{L}(t,\varepsilon)   &  	0	& 	0 \\
			0		&	0	& 	0 \\
			0		&	0	& \Sigma_{R}(t,\varepsilon)\\
		\end{array}
	\right].
\end{equation}

The self-energy \eqref{eq: selfenergyL} involves the inverse of a large-dimensional (or infinite-dimensional) matrix. Similar to the inversion in \eqref{eq: reduced-G},  it can be efficiently computed using a recursive algorithm, which has been well documented\cite{williams1982green,sancho1984quick,sancho1985highly}. The self-energy only needs to be computed once for constant external potential and for periodic external potentials, it can be pre-computed for one period.

Let {$\rho_{c}$} be the density-matrix restricted in the extended center region $\Omega_{\widetilde{C}}$, \ie,
\begin{equation}
\begin{aligned}
	\rho_c(t) := \rho(t)|_{\Omega_{\widetilde{C}}\times\Omega_{\widetilde{C}}} = D(t) + \rho_{c}(0).
\end{aligned}
\end{equation}
The reduced model for this part of the density-matrix can now be written as,
\begin{equation}\label{eq: reducedLvN}
	i \frac{d}{dt} \rho_c(t) = [H_\text{eff}(t), \rho_c(t)] + {\Theta_c}(t),
\end{equation}

With our choice of the subspaces, the reduced dynamics is  driven by the effective Hamiltonian $H_\text{eff}$. The non-homogeneous term 
 ${\Theta_c}$ embodies the effect of the potential, 
\begin{equation}\label{eq: theta'}
\begin{aligned}
	\Theta_{c}(t) = M^{*}\widetilde{V}^{*}\left(\varepsilon^{*}I-H\right)^{-1}\Theta(t) \left(\varepsilon I-H\right)^{-1}\widetilde{V}M,
\end{aligned}
\end{equation}
where $M$ is computed from Eq \eqref{eq: mass} and $\widetilde{V}$ is in the form of 
\begin{equation}
\widetilde{V} = 
	\left( 
		\begin{array}{ccc}
	         V_{L}      									        &	-H_{LC} 	     &       0  \\
		-H_{C\Gamma_{\!L}} V_{L}^{*}(\varepsilon-H_{LL})V_{L} 	&          \varepsilon - H_{CC} &    -H_{C\Gamma_{\!R}} V_{R}^{*}(\varepsilon-H_{RR})V_{R}  \\
		 0	             & 	     -H_{RC}		&         V_{R}
		\end{array}
	\right).
\end{equation}

The practical implementation of the reduced model hinges on the availability of efficient algorithms to compute  (i) the self-energy  \eqref{eq: selfenergyL}; (ii) the initial density-matrix in the center and contact region;  and (iii) the non-homogeneous term \eqref{eq: theta'}. The computation of the self-energy and the initial density-matrix,
as previously discussed, can be computed using the selective inversion techniques, which is applicable for problems that can be cast into the form of \eqref{eq: reduced-G} where the Green's function is accompanied by sparse vectors.  
As for the non-homogenous term, we find that $V_{\alpha}$ and $H_{\alpha,C}, \alpha=L,R$ have non-zeros elements only associated with those degrees of freedom in the domain $\widetilde{\Omega}$, which implies the sparsity of  $\widetilde{V}$. Upon closer inspection, we find that the product of inverse matrices in $\Theta(t)$, \ie, $\left(\varepsilon^{*}I-H\right)^{-1}\Theta(t) \left(\varepsilon I-H\right)^{-1}$, can be written as a sum of single matrix inverses (partial fractions), provided that $\varepsilon$ is in the resolvent of $H$ and $\text{Im}(\varepsilon)\ne 0$. For example, we have,
\[  \left(zI - H\right)^{-1}  \left(\varepsilon I-H\right)^{-1} =\frac1{\veps-z} \Big( \left(zI - H\right)^{-1}  - \left(\varepsilon I-H\right)^{-1}\Big).\]
Consequently, all those blocks can be written in the general form  \eqref{eq: reduced-G}, and one compute $\Theta_{c}$ efficiently by using the selective inversion techniques \cite{lin2011selinv}.

\subsection{Properties of the reduced models}
\subsubsection{ The Hermitian  property of $\rho_{c}(t)$}
The projection method produces an approximation of the density-matrix in the extended center region, leading to an open quantum-mechanical model that can be subsequently used to predict the current. 
The influence from the infinite leads, through the self-energy, has been implicitly incorporated into the effective Hamiltonian. 
By taking the Hermitian of the reduced model \eqref{eq: reducedLvN}, and noticing the anti-Hermitian property of the term $\widetilde\Theta$, we find that $\rho_{c}^{*}$ also satisfies \eqref{eq: reducedLvN} with initial condition $\rho_{c}^{*}(0)$. As $\rho_{c}(0)$ is Hermitian, and in light of the uniqueness of the solution, we obtain the Hermitian property for $\rho_{c}(t)$.

\subsubsection{The stability of the reduced models}
Next, let us turn to the analysis of stability. Since the stability of linear non-homogeneous system is implied by the stability of homogeneous system, we focus on the homogeneous case in Eq  \eqref{eq: reducedLvN} to study its stability. The problem can be addressed as the stability of a finite system $X(t)$,
\begin{equation}\label{eq: stability}
i\frac{d}{dt} X(t) = A(t)X(t) - X(t)A^{*}(t), \quad X(0) = \rho_{c}(0),
\end{equation}
where $A=H_{c}+\Sigma^{*}$. Since $\rho_{c}(0)$ has an eigen-decomposition $\rho_{c}(0) = \sum_{\ell} n_{\ell} \psi^{0}_{\ell}\psi^{0*}_{\ell}$, it is not difficult to verify that $X(t) = \sum_{\ell} n_{\ell}\psi_{\ell}(t)\psi_{\ell}^{*}(t)$ is the solution of Eq  \eqref{eq: stability} if $\psi_{\ell}(t)$ satisfies 
\begin{equation}\label{eq: waveequation}
	i\frac{d}{dt} \psi_{\ell}(t) = A(t)\psi_{\ell}(t), \quad \psi_{\ell}(0) = \psi_{\ell}^{0}.
\end{equation}
It suffices to analyze the stability of Eq  \eqref{eq: waveequation}.

There exists a decomposition $A(t)=A_{1}(t)+ i A_{2}(t)$, where $A_{1},A_{2}$ are real-valued symmetric matrices and $A_{2}$ is determined from $\Sigma$ due to the Hermitian property of $H_{c}$. Further computation yields,
\begin{equation}
A_{2}(t) = 
\left(
	\begin{array}{ccc}
	\widetilde{\Phi}_{L}\Lambda_{L}(t)\widetilde{\Phi}_{L}^{*} & 		0  		&     0  \\
		0	&     0  &  0  \\
		0     &     0  & \widetilde{\Phi}_{R}\Lambda_{R}(t)\widetilde{\Phi}_{R}^{*} \\
	\end{array}
\right),
\end{equation}
where $\widetilde{\Phi}_{\alpha} = H_{\Gamma_{\!\alpha},\alpha}\Phi_{\alpha}$ and $\Phi_{\alpha}$ is the eigenvectors of $H_{\alpha,\alpha}$. Thanks to the special form of $\Sigma$, one can compute that $\Lambda_{\alpha}$ is a real diagonal matrix, in the form, $\Lambda_{\alpha}=\text{diag}(\lambda^{\alpha}_{1},\lambda^{\alpha}_{2},\cdots,\lambda^{\alpha}_{n})$, with 
\begin{equation}
	\lambda^{\alpha}_{\ell} = \text{Im}\left(\frac{1}{\varepsilon^{*}-\mu^{\alpha}_{\ell}}\right),
\end{equation}
where $\mu^{\alpha}_{\ell}$ is the eigenvalue of $H_{{\alpha},{\alpha}}$. 

To ensure the stability, it is enough to require that $A_{2}$ has only non-positive eigenvalues \cite{brauer1966perturbations}, \ie, 
\begin{equation}
	\lambda^{\alpha}_{\ell} = \text{Im}\left(\frac{1}{\varepsilon^{*}-\mu^{\alpha}_{\ell}}\right) = \frac{\text{Im}(\varepsilon)}{|\varepsilon^{*}-\mu^{\alpha}_{\ell}|^{2}} \le 0.
\end{equation}
This confirms that when $\varepsilon$ has negative imaginary part, the stability of \eqref{eq: reducedLvN} is guaranteed.

\subsection{Higher order subspace projections}\label{sec: highorder}
The Galerkin-Petrov projection method can be extended to higher order, by expanding the subspaces $V_{L/R}$ and $W_{L/R}$ to higher dimensions. Here we provide two options to extend the current subspaces.

{\bf Expanding the contact region. } One straightforward approach is to keep the choices of $V$ and $W$ according to \eqref{eq: V'} and \eqref{eq: W'}, but increase the size of the region $\Omega_\Gamma$ to increase the subspace. Through numerical tests, we observe that this is a rather simple alternative, and it captures steady state current with subspaces of relatively small dimensions $n_\Gamma$. 

{\bf Block Krylov subspaces.} Another approach, as motivated by the block Krylov techniques \cite{ma2019coarse} for large-dimensional dynamical systems, is to expand the subspace $V_{L}$ to the block Krylov subspace,
\begin{equation}\label{eq: Krylov-V}
V_{L,m}=\left[ V_L \;\; H_{LL}V_L \; \cdots\; H_{LL}^{m-1}V_L \right] =: \mathcal{K}_{m}\left(H_{LL};V_{L} \right).
\end{equation}
The corresponding $W_{L,m}$ has a similar structure,
\begin{equation}\label{eq: Krylov-W}
	W_{L,m} = \left[ W_L \;\; V_L \; \cdots\; H_{LL}^{m-2}V_L \right] =: \mathcal{K}_{m}\left(H_{LL};W_{L} \right).
\end{equation}

The Krylov subspaces are composed of a generating matrix and a starting block.   In order to keep the additional blocks full rank, we pick $V_L$ based on the interaction range in $H_{LL}$. For example, if  $H_{LL}$ is based on a one-dimensional nearest-neighbor Hamiltonian, then we pick $n_\Gamma=1$ to define $V_L$, which would be a one-dimensional vector; We pick $n_\Gamma =2$ for a next nearest neighbor Hamiltonian, etc.  

\section{Numerical Experiments and Discussions}\label{sec: results}

To test the reduction method, we consider a one-dimensional two-lead molecular junction model within a TB setting. We follow the setup in   Zelovich et al. \cite{zelovich2014state}. More specifically, in the computation, the leads are represented by two finite atomic chains with increasing lengths ( $n_{L}$ and $n_{R}$ respectively) to mimic an infinite dimensional system and eliminate the finite size effect. The extended molecule with  length $n_{C}$ is represented by a finite atomic chain coupled with both leads. 
Here, the atomic unit is used throughout the paper if not stated otherwise.  

Initially, the system is configured in thermodynamic equilibrium, with all single-particle levels occupied up to the Fermi energy $\varepsilon_{F}=0.3$. The on-site energy is taken as $\alpha=2$, and the hopping integral between nearest neighbors is $\beta=-1$. At time $t = 0_+$, a bias potential is switched on in the electrodes. With the computed density-matrix, we study the bond current through the molecular junction to monitor the dynamics, using the formula \cite{zelovich2014state}
\begin{equation}
	I(t) = 2\beta \text{Im}[\rho_{j,j+1}(t)].
\end{equation}
For the time propagation of the density-matrix, we use the fourth-order Runge-Kutta scheme to solve the full model \eqref{eq: LvN}, as well as \eqref{eq: reducedLvN}. We fix the size of the center region $n_{C}=20$ and simulate the system under two different types of  external potentials: (1) constant biased potential: $U_{L/R}= \mp \frac{\delta U}2$ to mimic direct current (DC) circuit; (2) time-dependent potential: A sinusoidal signal in the left lead, $U_L=\sin \omega t,$ to mimic an alternating current (AC). 

\begin{figure}[H]
\includegraphics[scale=0.45]{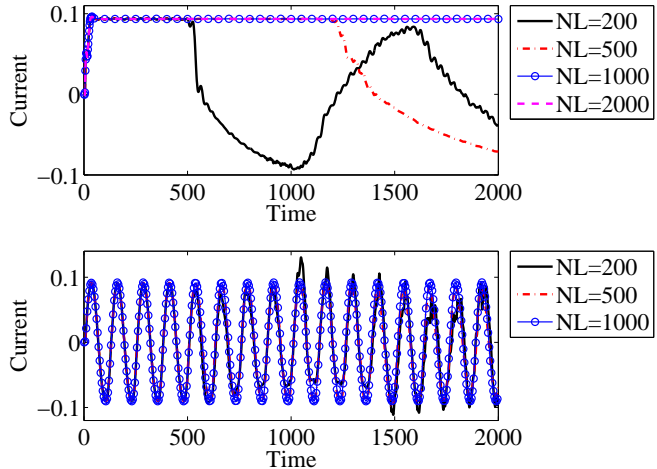}
\caption{(Color online) The finite size effect on the electric current. The figures show the time evolution of the currents through a junction coupled with leads of different lengths. Top: constant bias potential $U_{L} = -U_{R}=0.1$. Bottom: dynamic potential $U_{L}=0.2\sin(0.05t), U_{R}=0$.}
\label{fig: currentfull}
\end{figure}

In principle, the bath size needs to be infinite to model the two semi-infinite leads; but in  computations, one can only treat a system of finite-size  and  expect the system to reach a steady state in the limit as the bath size goes to infinity. First, we examine such size effect by varying $n_L$/$n_R$ and observing the current in the center  region.   More specifically, we run direct simulations using $n_{L}=n_{R}=200,500,1000,2000$. Our results (Figure \ref{fig: currentfull}) suggest that, for the constant potential case, the electric current  gradually develops into a steady state until the propagating electronic waves reach the ends of the leads and get reflected toward the bridge. As we extend the leads size to $n_{L}=n_{R}=1000$, the backscattering effect occurs much later and is no longer observed within the time window of our simulation. For the dynamic potential case, we observe periodic changes of the electric current. Size effects become insignificant when the size is increased to $n_{L}=n_{R}=500$ over the duration of the simulation. We point out that this effort of using sufficiently large bath size is only to generate a faithful result from  the full model \eqref{eq: LvN}, to examine the accuracy of the reduced model  \eqref{eq: reducedLvN}.

\begin{figure}[H]
\includegraphics[width=0.48\textwidth]{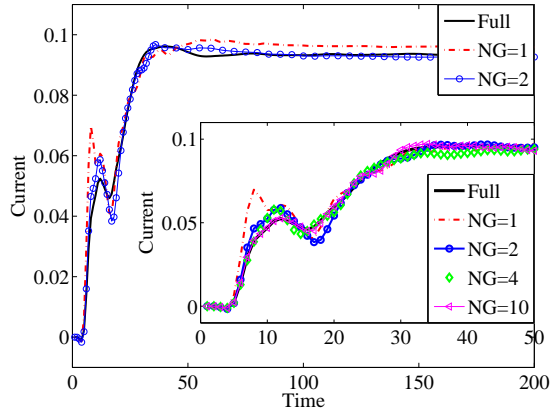}
\caption{(Color online) The simulation of the DC circuit (constant bias $U_{L}=-U_{R}=0.1$). The figure shows the time history of the current from the reduced model \eqref{eq: reducedLvN} with different subspace dimensions, compared to the result from the full model \eqref{eq: LvN}. The subspaces are chosen from \eqref{eq: V'} and \eqref{eq: W'} by extending the contact region $\Omega_{\Gamma}$ with parameter $\varepsilon=0.3-0.1i$. The inset shows the transient stage of the current.  }
\label{fig: DC}
\end{figure}

Next we compute the transient current of the DC circuit (case 1) from the effective reduced models  \eqref{eq: reducedLvN} and compare it with the current from the full model \eqref{eq: LvN} to evaluate the accuracy of the reduction method.  We also examine the  different choices of increasing the subspaces (as discussed in section \ref{sec: highorder}). In particular, in Figure \ref{fig: DC} we show the numerical results from using the subspaces \eqref{eq: V'}  and \eqref{eq: W'}, and we choose the dimension $n_\Gamma$ from 1 to 10.  First we notice that no recurrent phenomenon is observed, which can be attributed to the non-homogeneous term $\Theta(t)$ as well as the self-energy in Eq \eqref{eq: reducedLvN}, since they take into account the influence from the bath. The results improve as we expand the subspace, $V_{\alpha}$ and $W_{\alpha},\alpha=L/R$ in Eq \eqref{eq: VW}. The steady state current has already been well captured by the reduced model with  dimensions $n_\Gamma=2$, while the transient results improve as we expand  $n_{\Gamma}$, and we arrive at a very satisfactory result when $n_{\Gamma}=4$. 

We also tested the Krylov subspaces according to \eqref{eq: Krylov-V} and \eqref{eq: Krylov-W}. The subspaces can be expanded by increasing $m$. The steady state is well captured when $m=3,$ the transient requires higher order approximations.  Our observation is that in order to achieve the same accuracy, we need larger subspaces than the previous approach. On the other hand, the Krylov subspace approach is more robust in the regime where $\text{Im}(\varepsilon)$ is close to zero.
\begin{figure}[H]
\includegraphics[width=0.48\textwidth]{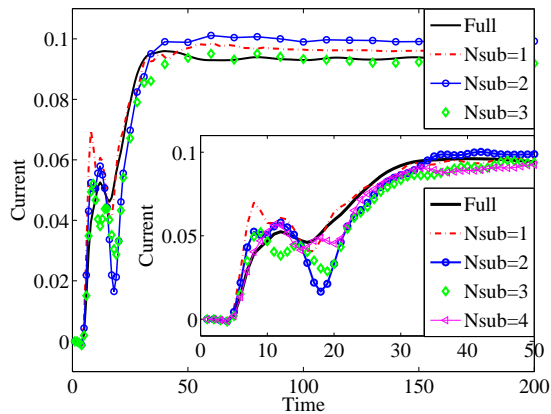}
\caption{(Color online) The results from the simulation of the DC circuit (constant bias $U_{L}=-U_{R}=0.1$). The figure shows the time evolution of the current from  the reduced model \eqref{eq: reducedLvN}, generated by the block Krylov subspaces \eqref{eq: Krylov-V} and \eqref{eq: Krylov-W} for various choices of dimensions (Nsub=$m$), with parameter $\varepsilon=0.3-0.01i$. The results are compared to the result from the full model \eqref{eq: LvN}. The inset shows the transient stage of the current.}
\label{fig: DC-krylov}
\end{figure}

\begin{figure}[H]
\includegraphics[height=0.33\textwidth]{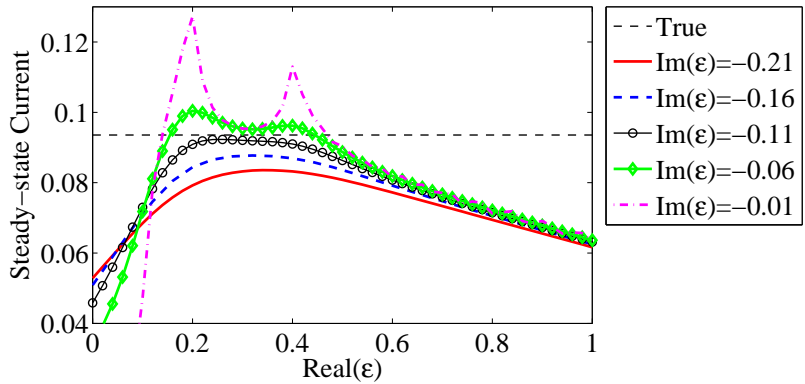}
\caption{(Color online) The example of DC circuit with constant bias ($U_{L}=-U_{R}=0.1$). The Figure shows the  steady-state current predicted by the  reduced model \eqref{eq: reducedLvN}  using various choices of the parameter $\varepsilon$ with $n_{\Gamma}=1$.}
\label{fig: DC'}
\end{figure}

Another important factor that plays a role in the reduced model is the selection of the parameter $\varepsilon$, which can be viewed as an interpolation point for the self-energy.  Therefore, we study the dependence of $\varepsilon$ in the reduced models, by observing the electric current at steady state for various different choices of $\veps$. For the imaginary part, we require $\text{Im}(\veps)$ to be strictly less than zero to ensure that the self-energy \eqref{eq: selfenergyL} is well defined and \eqref{eq: reducedLvN} has the stability assurance. We start with  $\text{Im}(\veps)=0.1$.  
 When $|\text{Im}(\varepsilon)|$  is further decreased ($<0.01$), the electric current exhibits oscillations around the true value of the steady state. For the real part of $\veps$, the optimal value  appears around the Fermi energy.  See Figure \ref{fig: DC'}. This suggests that $\veps$ should be around the Fermi level with small imaginary part, although when the imaginary part is too small,  the numerical robustness might be affected.
 
\begin{figure}[H]
\includegraphics[width=0.48\textwidth]{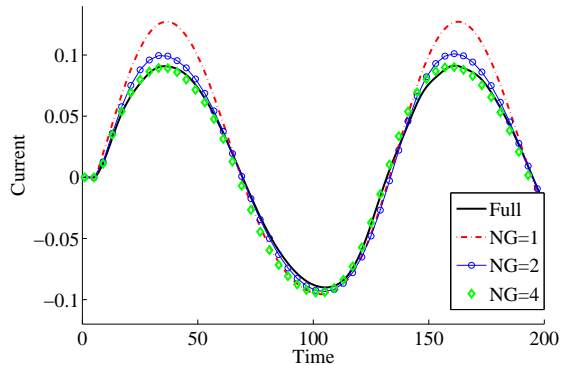}
\caption{(Color online) The example of an  AC circuit with time-dependent potential $U_{L}(t)=0.2\sin(0.05t), U_{R}=0$. This figure displays the time evolution of the currents from the  reduced models with different subspace dimensions $n_{\Gamma}$, compared to that from the full model. The parameter $\varepsilon = 0.3-0.1i$ is used.}
\label{fig: AC}
\end{figure}

Finally, we turn to the example of the AC circuit. Since a time-dependent external potential is imposed, $H_{c}$ and $\Theta_{c}$ in Eq \eqref{eq: reducedLvN} are time-dependent  as well.  They need to be evaluated at each time step. Due to the periodic property, it suffices to pre-compute $H_{c}(t)$ and $\Theta_{c}(t)$ within one time period. As shown in Figure \ref{fig: AC}, a periodic electric current has been reproduced by the reduced model  \eqref{eq: reducedLvN},  and the accuracy also improves as we expand the subspace size $n_{\Gamma}$. The electric current is already well captured when $n_{\Gamma}=4$. 

\section{Summary}\label{sec: summary}
We have proposed to formulate the quantum transport problem in a molecular junction coupled with infinite baths as a reduced-order modeling problem. The goal is to derive a finite quantum system with open boundary conditions. 
Motivated by the works \cite{sanchez2006molecular,zelovich2014state,zelovich2015molecule}, we work with the density-matrix, and obtain reduced Liouville-von Neumann  equations 
for the center and contact regions. The reduced equations are derived using a systematic projection formalism, together with appropriate choices of the subspaces. 
Numerical experiments have shown that the reduced model is very effective in capturing the steady-state electric current as well as the transient process of the electric current. 
The accuracy increases as we expand the contact regions in the reduced model. 

In order to demonstrate the reduction procedure,  we have considered a one-dimensional junction system. But the validity of the projection approach is not restricted to the one-dimensional system. It can be applied to general coupled system-bath dynamics that require model reduction due to the computational complexity.  The extension to systems that are of direct practical interest is underway. Another possible extension is the data-driven implementation of reduced-order modeling. In this case, rather than computing the  matrices in the reduced models from the underlying quantum mechanical models, they are inferred from observations \cite{benner2015survey,ma2019coarse}.

Self-consistency has not been included in the Liouville-von Neumann  equation, especially the Coulomb potential, which in the linear response regime, leads to a dense matrix \cite{yabana2006real} from the Hartree term. This creates considerable difficulty for the reduce-order modeling since  the partition \eqref{eq: tildeH} is no longer reasonable.  However, the Coulomb and exchange correlation are known to be important for the Coulomb blockade phenomena \cite{kurth2010dynamical}. This difficulty in the modeling of quantum transport has also been pointed out  in \cite{kurth2005time,ullrich2011time}. In practice, this is often dealt with by solving Poisson's equation in a relatively larger domain with Dirichlet boundary conditions \cite{taylor2001ab}. We will address this issue under the framework of reduced-order modeling in separate works.

\begin{acknowledgement}
This research was supported by NSF under grant DMS-1619661 and DMS-1819011.

\end{acknowledgement}

\bibliography{JCTC}

\end{document}